      [1999/12/01 v1.4c Il Nuovo Cimento]
\documentclass{cimento}

\usepackage{graphicx}  
\title{Can we use GRB to probe high redshift star-formation?}
\author{F.~Fiore\from{ins:x}\ETC,
D.~Guetta\from{ins:x},
S.~Piranomonte\from{ins:x},
V.~D'Elia\from{ins:x},\atque
L.A.~Antonelli\from{ins:x}
\instlist{\inst{ins:x} INAF-OAR - Monteporzio, Italy}}

\PACSes{
\PACSit{95.55.Ka}{cosmology-observations}
\PACSit{98.70.Rz}{$\gamma$-ray sources; $\gamma$-ray bursts}
} 

\def\ls{{_<\atop^{\sim}}}

\begin{document}

\maketitle

\begin{abstract}
We present a detailed analysis of the selection effects
that plague GRB observations. We find that these effects may
partially explain the different redshift distributions between 
BeppoSax/HETE2 and Swift bursts. It is mandatory to consider 
these effects to determine the redshift evolution of GRBs.

\end{abstract}

\section{Introduction}

Gamma Ray Bursts (GRBs) are one of the great wonders of Universe. They
combine several of the hottest topics of 21$^{st}$ century
astrophysics. On one side they are privileged laboratories for
fundamental physics, including relativistic physics, acceleration
processes and radiation mechanisms. On the other side, being GRBs
associated to the death of massive stars, they can be used as 
cosmological tools to investigate star-formation and metal enrichment
at the epochs of galaxy birth, formation and growth. There are two
main research areas to the latter purpose.  The first one uses the GRB
as a background beacon for spectroscopy of UV lines, to characterize
the physical and chemical status of the matter along the line of
sight.  The second includes statistical studies of the GRB redshift
distributions \cite{ref:guetta05} to infer information on the
histories of the GRB- and star- formation rates and of the metal
enrichment in the Universe (e.g. \cite{ref:pm2001}).

Although the techniques adopted, and therefore the reference
communities, are somewhat different, these research areas are 
interconnected. As an example, on one hand spectroscopy of UV lines
can be used to determine the metal content of the absorption
systems. On the other hand, it is well known that galaxy scale
properties like metalicity, star-formation rate and mass are
correlated. Then, metalicity determinations obtained through GRB
spectroscopy could be, at least in principle, plugged in the GRB
population studies, to obtain a better constraint on the models.

Unfortunately, both research areas are plagued by large uncertainties.

(1) Absorption line spectroscopy is certainly a powerful
tool. Spectroscopy of galaxies and QSOs is routinely used to study
both the galaxy Inter-Stellar-Matter (ISM) and the
Inter-Galactic-Matter (IGM) along the line of sight. However,
star-formation regions and the ISM in general, are highly complex
systems. It is then not clear whether single lines of sight can
provide information truly representative of such systems as a whole.

(2) Population studies are also very powerful tools. Galaxy and AGN
counts and luminosity functions have been used to successfully measure
the evolution of the star-formation rate, of the galaxy and black hole
mass densities up to z=3-4 and beyond. Thanks to first BeppoSAX and
then to HETE2 and Swift we start to have sizable samples of reliable
GRB redshifts (about 70 up to now), and a similar number will likely
be produced by Swift in the next few years. This opens up the
possibility to compute relatively well constrained GRB luminosity
functions in a few redshift bins, and therefore measure the cosmic
evolution of the GRB rate. However, the fraction of Swift GRBs with a
reliable redshift is today about one third of the total. It is
expected that this fraction will improve in future, but it will hardly
approach to the majority of the GRBs. This means that the biggest
problem we have to face to use GRBs as cosmological tools is to
understand and account for large selection effects.

In the next sections we address the second problems while the first is
discussed in a companion paper \cite{ref:ff2006}.

\section{GRB population studies}

\begin{figure}
\begin{center}
\includegraphics[height=8truecm,width=8truecm]{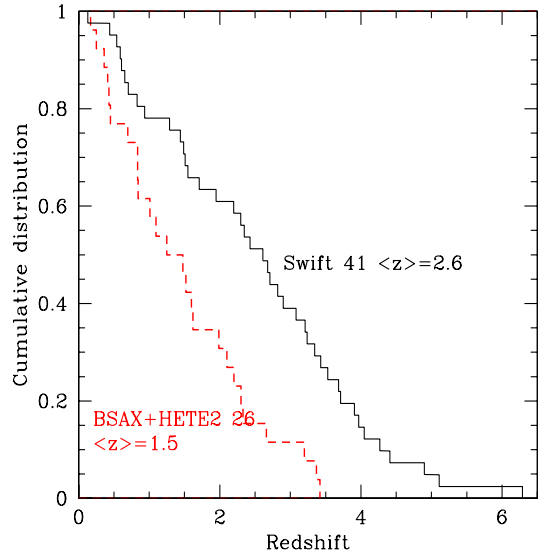}
\caption{Cumulative redshift distributions of Swift (solid line) 
and BeppoSAX+HETE2 (dashed line) GRBs.}
\label{zdist}
\end{center}
\end{figure}

The importance of large selection effects shaping the population of
GRBs with a measured redshift is evident when comparing the redshift
distribution of Swift GRBs with that of BeppoSAX and HETE2 GRBs
(figure \ref{zdist}). The median redshifts of the two distributions are
2.6 and 1.5 respectively.  
This discrepancy can not be simply explained as due to a different 
detector sensitivity (Guetta \& Piran in preparation). 
To gain more information on this issue we
study three well defined sample of GRBs detected by these three
satellites. We select GRBs outside the Galactic plane, to limit
Galactic extinction along their line of sight and to avoid fields too
crowded, which can complicate the discovery of optical/NIR afterglows,
so hampering redshift determinations. To this purpose we limit our
study to regions with Galactic column density along the line of sight
$<2\times10^{21}$ cm$^{-2}$. We also select GRBs with good (arcmin)
localization.  For BeppoSAX and HETE2 GRBs we require that the
$\gamma-$ray burst is detected by the high energy GRBM and Fregate
instruments and is localized by the WFC, WXC or SXC instruments. For
Swift we consider all GRBs up to September 10 2006. We consider only
reliable spectroscopic redshift. Table 1 gives more information on the
selected samples.

\begin{table}
\caption{GRB samples: the ones with optical afterglows (O.A), the ones with
optical afterglow temporal decay index (O.decay), the total number of GRBs 
with z and the ones with z derived from the host galaxy}
\begin{tabular}{lccccc}
\hline
Sat.      & Tot. GRB & O.A. & O. decay & Tot. z spec. & z from host gal. \\
\hline
Swift     & 122      & 62 & 44    &   41        &  6 \\
HETE2	  & 44       & 17 & 15    &   14        &  2 \\
BeppoSAX  & 39       & 18 & 16    &   12        &  3 \\
\hline
\end{tabular}
\end{table}

There are two large groups of selection effects that must be
considered at least: (1) GRB detection and localization and (2)
redshift determination through spectroscopy of the optical/NIR
afterglow or of the GRB host galaxy. We discuss these two issues in
the next sections.

\subsection {GRB detection and localization}

The sensitivity of BeppoSAX, HETE2 and Swift instruments as a function
of the GRB spectral shape has been studied in detail by
\cite{ref:b2003,ref:b2006}.\cite{ref:b2006} also studied the sensitivity 
of the BAT instrument as a function of the combined GRB temporal and
spectral properties. We refer to these papers for more details on
these topics.

\begin{figure}
\begin{tabular}{cc}
\includegraphics[height=6.8truecm,width=6.8truecm]{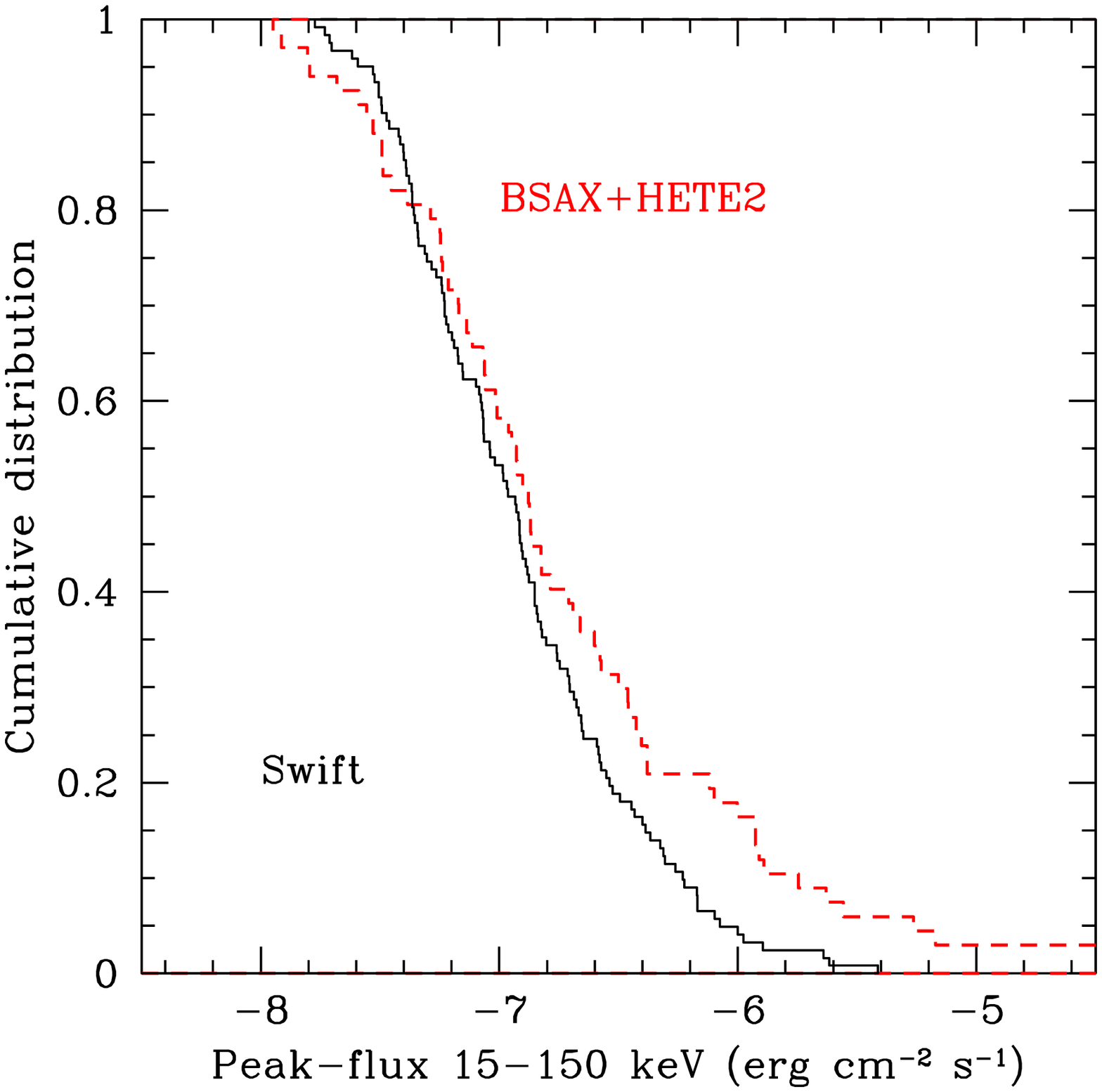}
\includegraphics[height=6.8truecm,width=6.8truecm]{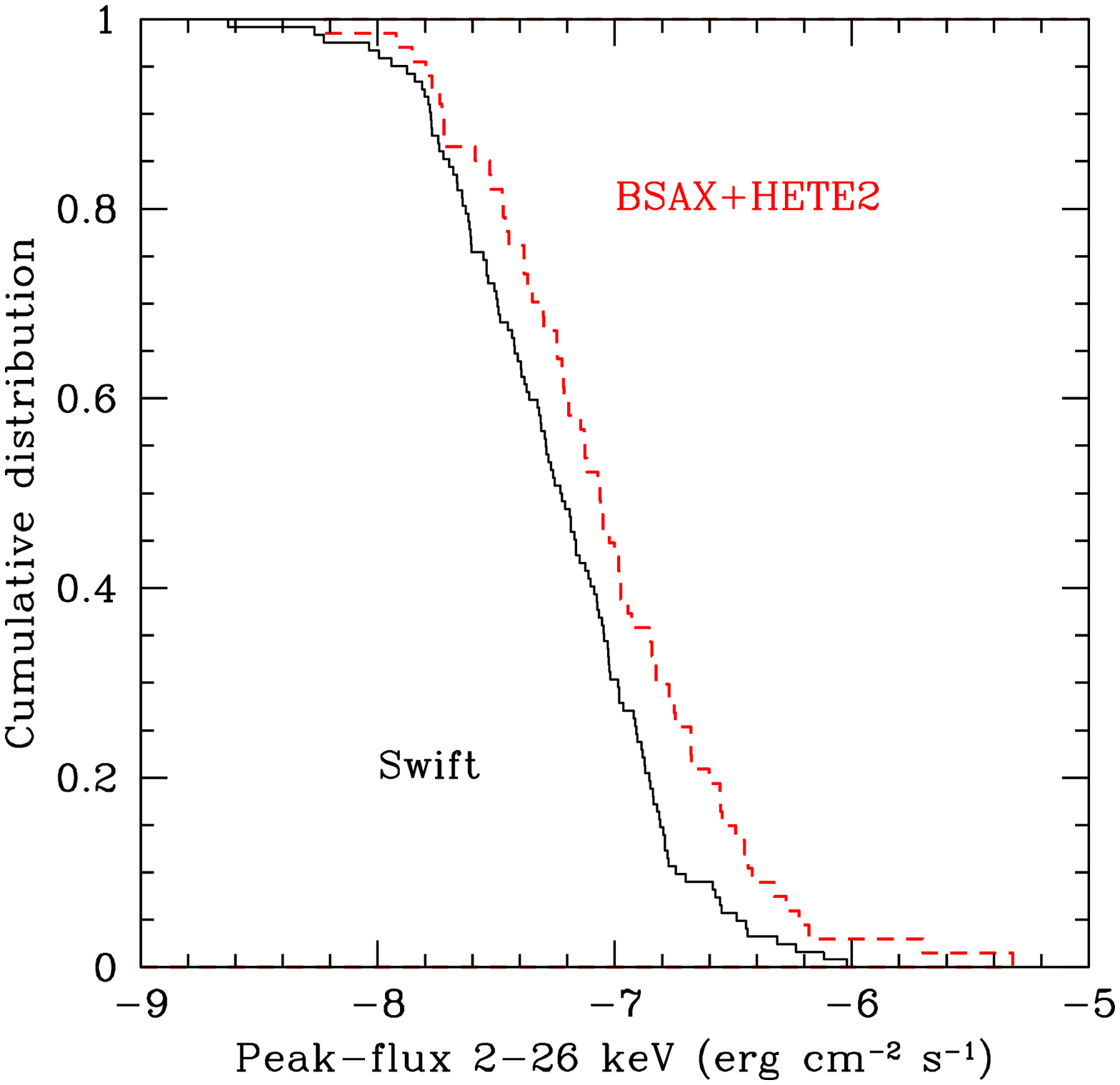}
\end{tabular}
\caption{Peakflux cumulative distributions of the Swift (solid line) 
and BeppoSAX+HETE2 (dashed line) GRBs.
a), left panel, 15-150 keV band; b), right panel, 2-26 keV band.}
\label{pfdist}
\end{figure}

Figure \ref{pfdist} compares the peak-flux cumulative distributions of
the Swift GRBs with that of BeppoSAX and HETE2. The comparison is done
in two energy bands: 15-150 keV, which is the band where BAT detects
and localizes GRBs, and 2-26 keV, which is the band where the BeppoSAX
WFC and the HETE2 WXC and SXC localize GRBs. To produce figure
\ref{pfdist}a) GRBM and Fregate peak fluxes were converted to the
15-150 keV BAT band using a power law model with an average energy
index of 0.5 for the BeppoSAX burts and the best fit model in
\cite{ref:sak2005} for the HETE2 bursts. BAT peak-fluxes and spectral
parameters are taken from the Swift GRB Information page\footnote{
http://swift.gsfc.nasa.gov/docs/swift/archive/grb\_table.html}.  To
produce figure \ref{pfdist}b) we used WFC and WXC peak-fluxes and
converted BAT 15-150 keV peak fluxes in the 2-26 keV band using the
best fit models and parameters and the best fit observed column
densities along the line of sight to the GRBs. To assess the
robustness of our analysis we produced peak-flux cumulative
distributions using different, but reasonable, values of the spectral
parameters adopted for the conversion from one band to the other. We
always found results qualitatively similar to those in figure
\ref{pfdist}.

Figure \ref{pfdist}a) shows that Swift is finding, on average, GRBs
slightly fainter than BeppoSAX and HETE2 in the 15-150 keV band. The
BeppoSAX and HETE2 samples contain a higher fraction of bright
GRBs. The median log(peak-flux) and its interquartile range are -6.93,
0.33 for the Swift sample and -6.88, 0.42 for the joined
BeppoSAX+HETE2 sample. This is expected because of the better
sensitivity of the BAT instrument with respect to the BeppoSAX GRBM
and HETE2 Fregate instruments (\cite{b2003}).
The median 2-26 keV log(peak-flux) is -7.22, 0.32 for the Swift sample
and -7.06, 0.38 for the joined BeppoSAX+HETE2 sample. The two
2-26 keV distributions differ more than the 15-150 keV
peak-flux distributions.  This is probably due to the fact that Swift
GRBs are localized at energies higher than 10-15 keV, while BeppoSAX
and HETE2 GRBs are localized at energies $\ls10$ keV. This implies
that Swift localizes, on average, GRBs harder and/or less affected by
photoelectric absorption than BeppoSAX and HETE2. In particular, the
latter GRB samples would be biased against flat spectrum GRBs and also
against large obscuration. A column density of N$_H=10^{23}$ cm$^{-2}$
at z=1 would reduce the observed 2-10 keV flux by 12-15\% (depending
on the spectral index), thus reducing the probability of a detecting
such highly obscured GRBs with BeppoSAX WFC and HETE2 WXC. Conversely,
they would certainly be present in the Swift sample.

\begin{figure}
\begin{tabular}{cc}
\includegraphics[height=6.8truecm,width=6.8truecm]{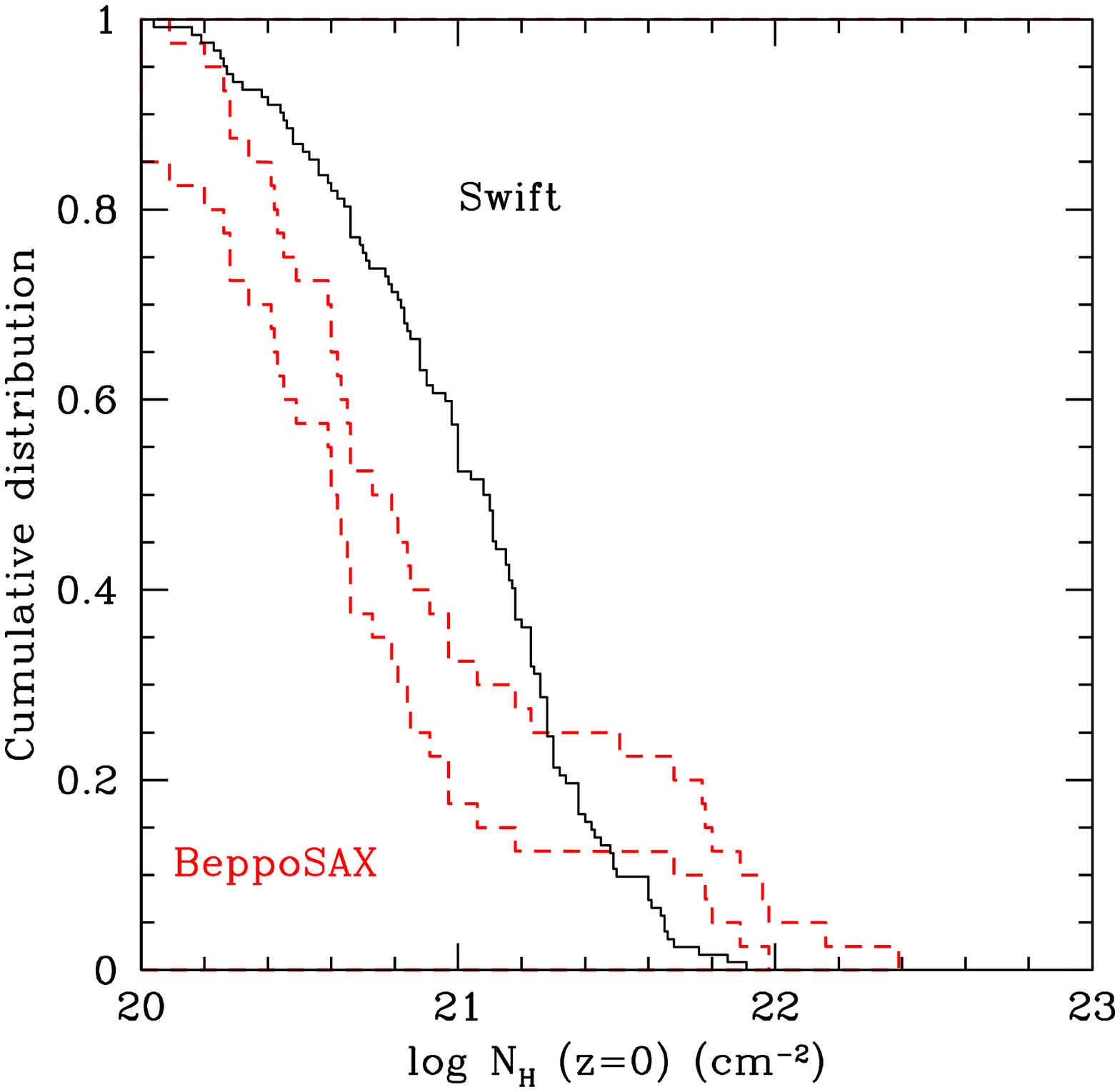}
\includegraphics[height=6.8truecm,width=6.8truecm]{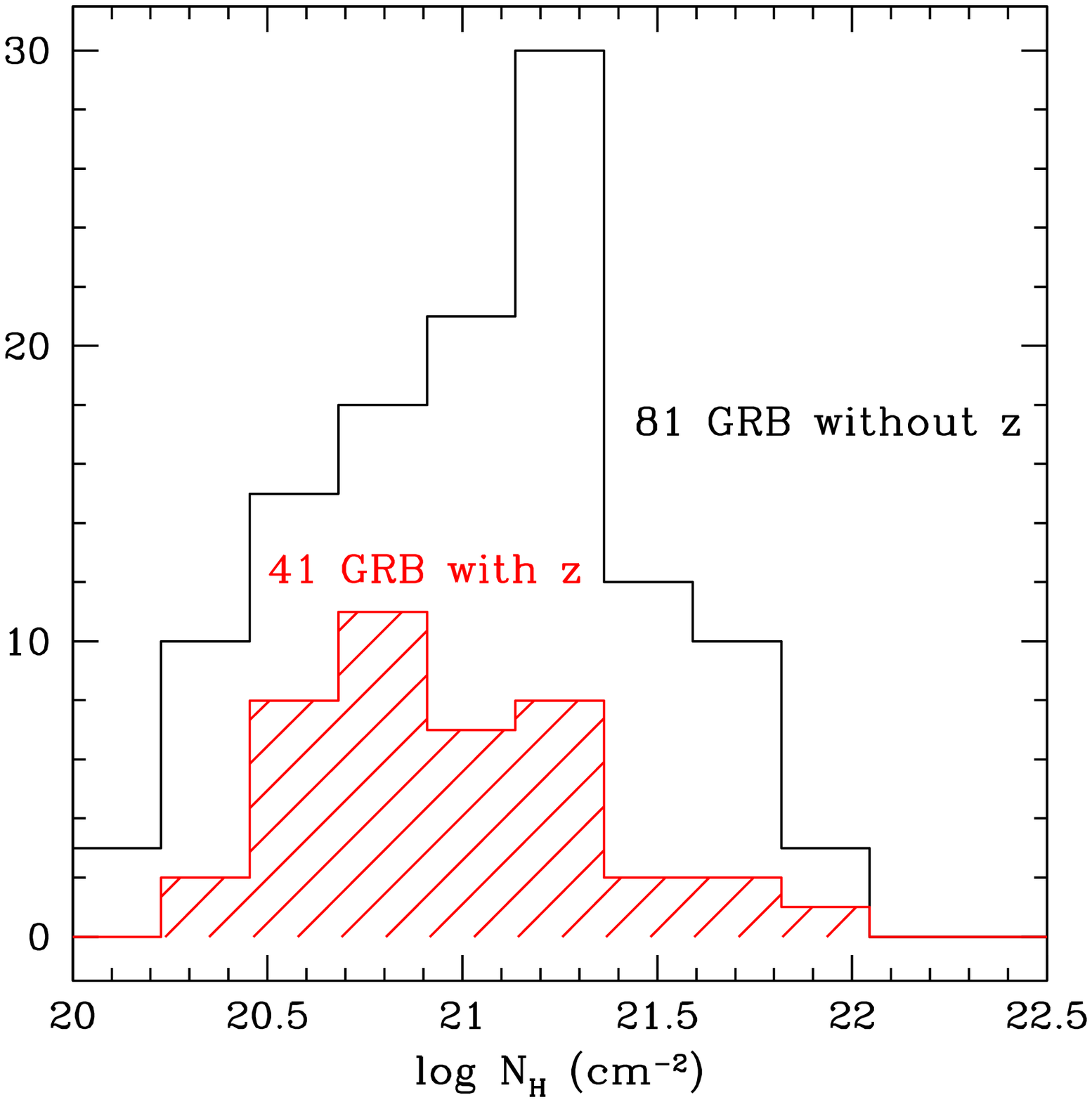}
\end{tabular}
\caption{a), left panel, N$_H$ cumulative distributions of the Swift 
(solid line) and BeppoSAX (dashed lines) GRBs.  The leftmost BeppoSAX
curve assumes N$_H$=N$_H{Galactic}$ for the GRBs with a best fit
intrinsic N$_H$ consistent with zero. The rightmost curve assumes for
these GRBs the 90\% upper limit. b), right panel, N$_H$ histograms of
the Swift GRB with (shadow histogram) and without redshift (black
histogram).}
\label{nhdist}
\end{figure}

Figure \ref{nhdist}a) compares the observer frame best fit column
density N$_H$ for the samples of Swift and BeppoSAX GRBs.  The
uncertainties on the X-ray spectral parameters, and therefore on
N$_H$, are much bigger for the BeppoSAX GRBs than for the Swift
GRBs. Indeed, the typical uncertainty on Swift column densities is
$5-10\times10^{20}$ cm$^{-2}$ (see e.g. \cite{ref:cam2006}), while
that on BeppoSAX column densities is $\approx10$ times higher (see
\cite{ref:st2004,ref:dep2003,ref:piro2005}).  For this reason we plot
2 curves for the BeppoSAX GRBs. The leftmost curve assumes
N$_H$=N$_H{Galactic}$ for the GRBs with a best fit intrinsic N$_H$
consistent with zero. The rightmost curve assumes for these GRBs the
90\% upper limit. The real BeppoSAX N$_H$ distribution is probabily
between the two curves.  Both BeppoSAX curves in figure \ref{nhdist}a)
are significantly lower than the Swift curve (probability $<10^{-5}$
and 1.7 \% respectively, using the Kolmogorov-Smirnov test), thus
confirming that Swift samples are less biased against obscuration than
the BeppoSAX sample. Since the observer frame column density scales as
the rest frame column density times (1+z)$^{-2.5}$, the BeppoSAX
sample would probably be biased against low-z, highly obscured
GRBs. Conversely, these GRBs must be present in the Swift sample.
However the determination of their redshift may be complicated by dust
extinction, which makes fainter the optical afterglow, if dust is
associated to the X-ray absorbing gas.  Indeed, the N$_H$ distribution
of the Swift GRB with a redshift is shifted toward lower N$_H$ values
than the distribution of GRB without redshifts (figure
\ref{nhdist}b). This introduces the next important group of selection
effects, those concerning the determination of the redshift of a GRB
through spectroscopy of the optical/NIR afterglow or of its host
galaxy.

\subsection{Redshift determination}

In determining the redshift of a GRB the identification of the optical
afterglow plays a major role. Only 6 redshift of the Swift GRB sample
considered here have been found through spectroscopy of the host
galaxy (5 for the BeppoSAX and HETE2 joined sample).

Optical afterglows have been discovered for 50\% of the Swift GRB
sample, a fraction only slightly greater than that of the joined
BeppoSAX and HETE2 sample (42\%). This result is somewhat surprising,
because the prompt Swift localization (minutes) and the large
international effort on Swift GRB follow-up, which can exploit an
impressive number of facilities, from dedicated robotic telescopes to
8m class telescopes like the VLT, Gemini and Keck, was thought to
produce a much larger fraction of optical/NIR afterglow identifications
than BeppoSAX and HETE2.

\begin{figure}
\begin{tabular}{cc}
\includegraphics[height=6.8truecm,width=6.8truecm]{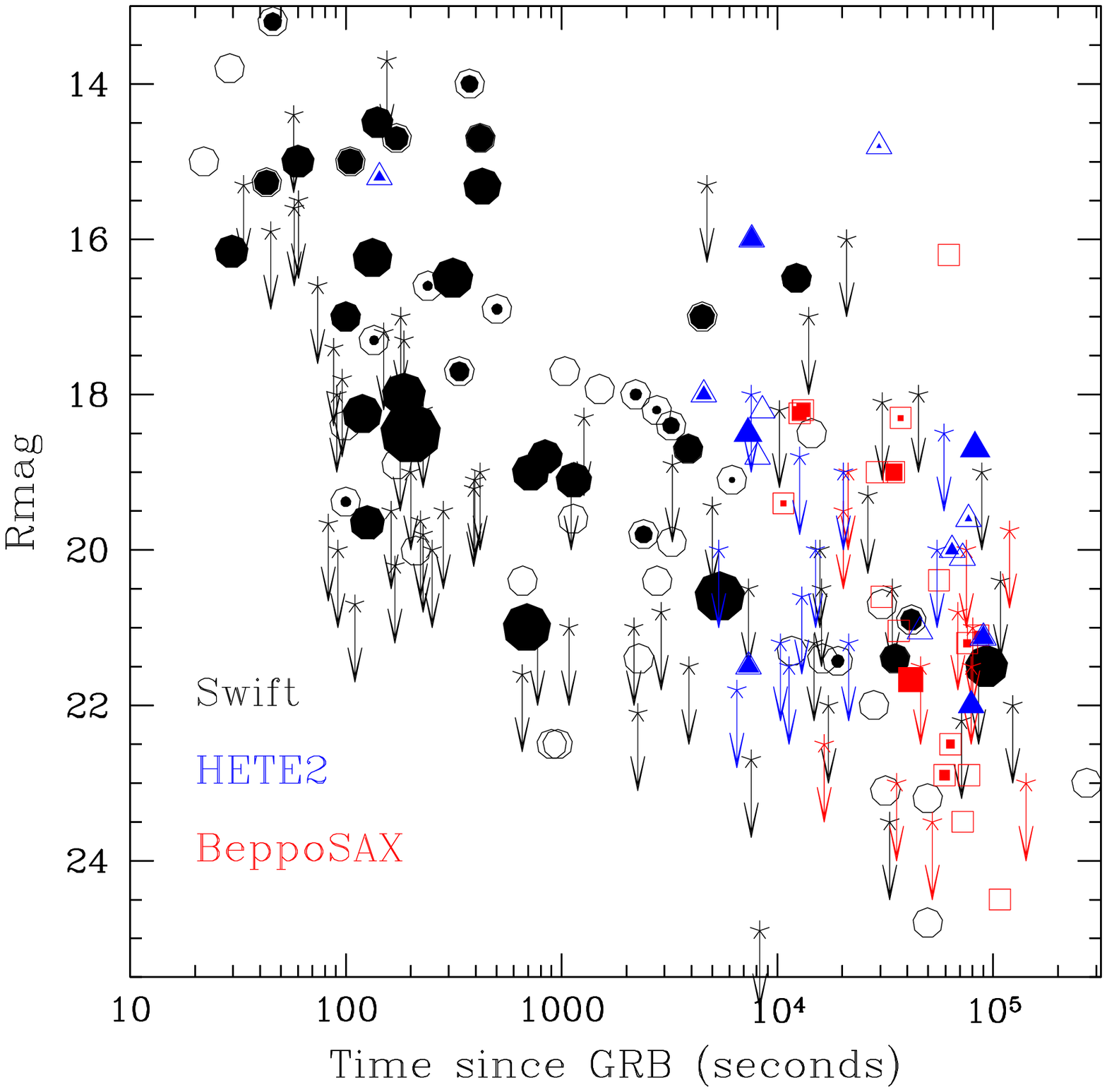}
\includegraphics[height=6.8truecm,width=6.8truecm]{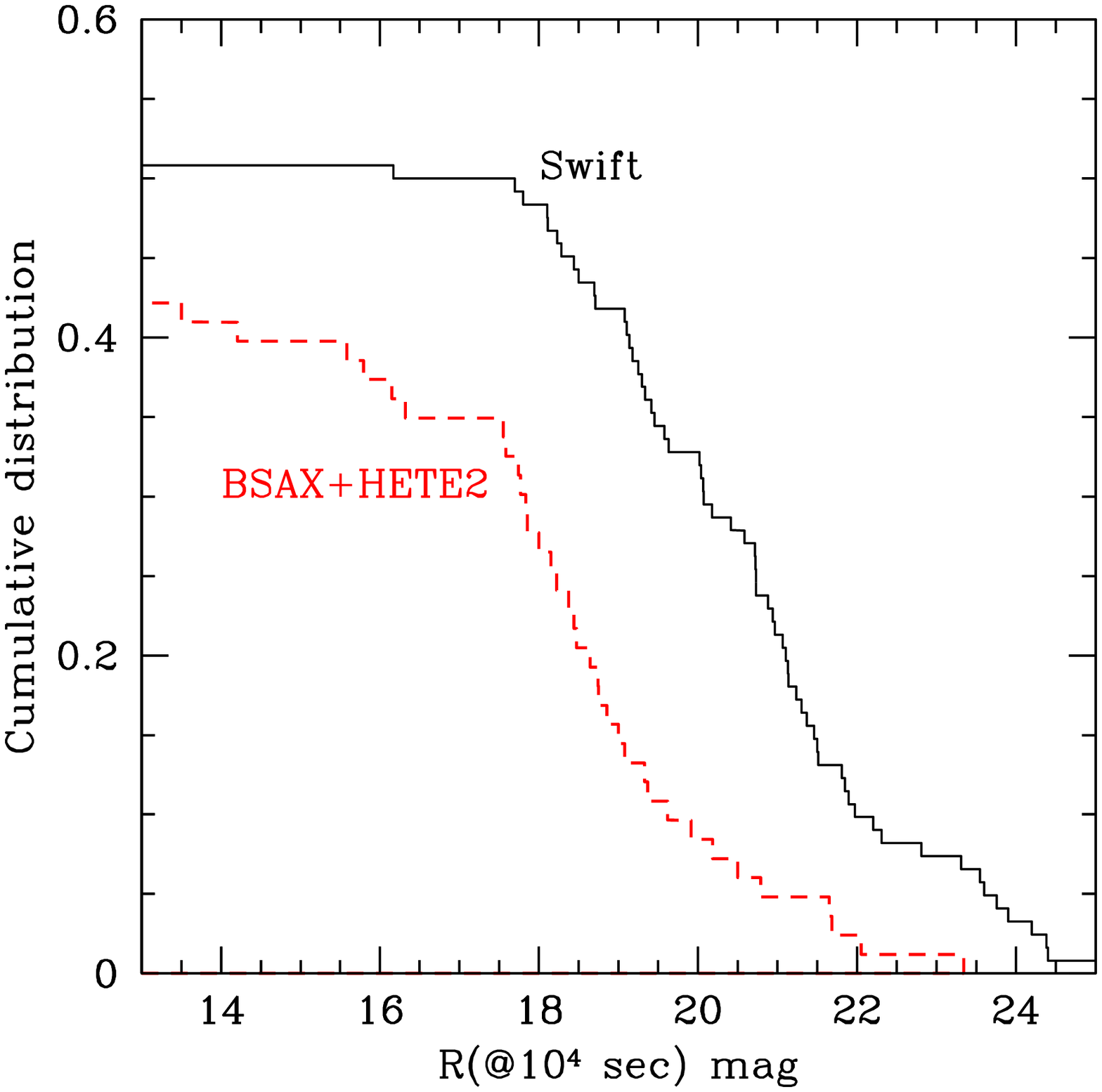}
\end{tabular}
\caption{a), left panel, the R magnitude of the optical afterglow at the time 
of its discovery as a function of this time. Filled symbols are GRB
with reliable redshift determination. The size of the symbol is
proportional to the redshift.  Circles = Swift GRBs; squares =
BeppoSAX GRBs; triangles = HETE2 GRBs.  b), right panel, the
cumulative distribution of the R magnitude 10ksec after the GRB events
for the Swift (solid line) and the BeppoSAX+HETE2 (dashed line) GRBs.}
\label{r3}
\end{figure}

Figure \ref{r3}a) shows the R magnitude of the optical afterglow at
the time of its discovery as a function of this time for the Swift GRBs
and the BeppoSAX and HETE2 GRBs. As expected redshifts are preferably
found for bright afterglows.  The figure also suggests that on
average and at each given time from the GRB even the Swift optical
afterglows are fainter than the BeppoSAX and HETE2 afterglows. We then
computed the magnitude of the Swift, BeppoSAX and HETE2 afterglows at
a fixed time using the best fit decay indices found for each GRB
afterglow, if available. In the rest of the cases we used a
decay index of -1. We chose a fixed time of 10 ksec after the burst
(observer frame).  Figure \ref{r3}b) compares the Swift distribution
of the R magnitude @ 10 ksec with that of the BeppoSAX and HETE2.
As we can see from this figure Swift detect much fainter bursts in the
optical band.

\section{Selection effects at works}

To understand how the different selection effects (peak-flux limit,
X-ray obscuration and magnitude of the optical afterglow) can modify
the redshift distribution we extracted from the Swift GRB sample a
subsample having the same peak-flux, N$_H$ and Rmag(@10ksec)
distributions of the joined BeppoSAX+HETE2 samples (the simulated
redshift distribution hereafter).  Figure 5 compares the
redshift distrubution of this subsample with the full Swift and
BeppoSAX+HETE2 GRBs. To evaluate the uncertainty on the simulated
redshift distribution we ran the random extraction 100 times.  We see
that the average simulated redshift distribution is consistent, within
the statistical uncertainties, with the real BeppoSAX+HETE2 redshift
distribution at all redshifts but z=1.5-2.  In particular, the
shallower peak-flux limit and brighter magnitudes of the optical
afterglow of the BeppoSAX+HETE2 GRBs explain why only few high
redshift GRBs are found in this sample. On the other hand, the smaller
number of low-z GRBs can be explained by the lower absorbing column
densities of the BeppoSAX GRBs.  It remains to understand the deficit
of z=1.5-2 GRBs in the BeppoSAX+HETE2 sample.  This can hardly be
explained by the three quantities considered in our analysis. A
possible explanation for this deficit can be related to the lack of
strong absorption features in the 3800-8000 \AA\ range covered by
optical spectrometers for this redshift interval, coupled with worse
quality spectroscopy available at least at the time of the BeppoSAX
measurements.  In fact, the Lyman$\alpha$ enters the above range only
at z$\sim2.3$, while the MgII doublet goes in a region strongly
affected by telluric features at z$>1.5$.  It is however difficult to
make more quantitative this issue.  

In conclusion, in order to safely use GRBs as cosmological tools and derive
their redshift evolution from statistical analysis of the present GRB
samples selection effects on GRB detection, GRB
localization and GRB redshift determination must be taken into account
properly (Guetta et al. 2006 in preparation).

\begin{figure}
\label{redshift}
\begin{center}
\includegraphics[height=8truecm,width=8truecm]{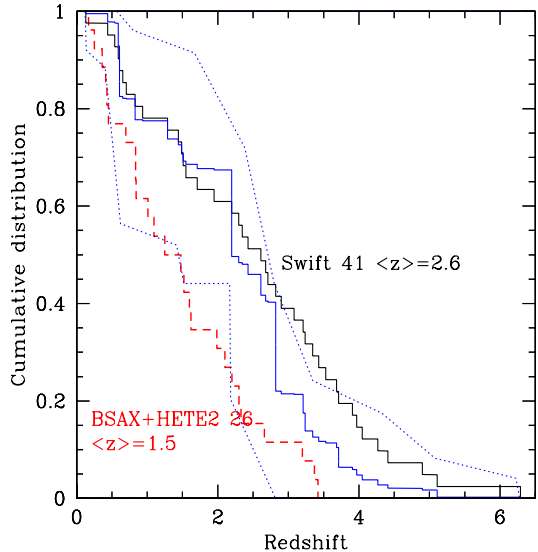}
\caption{The average cumulative redshift distribution of a subsample of 
Swift GRBs having the same peak-flux, N$_H$ and Rmag(@10ksec)
distributions of the joined BeppoSAX+HETE2 samples (thin solid line)
compared with the Swift (thick solid line) and BeppoSAX+HETE2 (dashed line) 
total redshift distributions. The dotted lines mark
the redshift range covered by 100 random extractions and can give an
idea of the statistical uncertainty associated to a single extraction
of a redshift distribution of 25 GRBs from a parent population.}
\end{center}
\end{figure}

\acknowledgments
We thank Eli Waxman for his useful comments.
We acknowledge support from contracts ASI/I/R/039/04 and  ASI/I/R/023/05/0.

\end{document}